\documentclass[12pt,a4paper]{article}

\usepackage{amsmath}
\usepackage{amssymb}
\usepackage{epsfig}
\usepackage{graphicx}
\usepackage{cite}
\newcommand{\capdef}{}
\newcommand{\mycaption}[2][\capdef]{\renewcommand{\capdef}{#2}%
       \caption[#1]{{\footnotesize #2}}}
\makeatletter
\renewcommand{\fnum@table}{\textbf{\tablename~\thetable}}
\renewcommand{\fnum@figure}{\textbf{\figurename~\thefigure}}
\makeatother

\newcounter{myenumi}

\renewcommand{\themyenumi}{\roman{myenumi}}
{\end{list}}

\newlength{\myem}
\newcounter{mysubequation}[equation]

\makeatletter
\renewcommand{\section}{\@startsection{section}{1}{0em}{-\baselineskip}%
{\baselineskip}{\normalfont\large\bfseries}}
\renewcommand{\subsection}%
{\@startsection{subsection}{2}{0em}{-0.7\baselineskip}%
{0.7\baselineskip}{\normalfont\bfseries}}
\makeatother

\newcommand{\ie}{{\it i.e.}}

\newcommand{\eg}{{\it e.g.}}

\newcommand{\cf}{{\it cf.}}
\newcommand{\etc}{{\it etc.}}
\newcommand{\eq}{Eq.}

\newcommand{\fig}{Fig.}

\newcommand{\Ref}{Ref.}
\newcommand{\Refs}{Refs.}

\newcommand{\Tab}{Table}

\newcommand{\bi}{\begin{itemize}}
\newcommand{\ei}{\end{itemize}}

\newcommand{\be}{\begin{equation}}
\newcommand{\ee}{\end{equation}}
\newcommand{\bea}{\begin{eqnarray}}
\newcommand{\eea}{\end{eqnarray}}

\newcommand{\ldm}{\Delta m_{31}^2}
\newcommand{\sdm}{\Delta m_{21}^2}
\newcommand{\deltacp}{\delta_{\mathrm{CP}}}
\newcommand{\stheta}{\sin^2 2 \theta_{13}}

\newcommand{\GLOBES}{{\sf GLoBES}}
\newcommand{\GLOBESN}{{\sf GLoBES~3.0}}
\newcommand{\AEDL}{{\sf AEDL}}

\newcommand{\equ}[1]{\eq~(\ref{equ:#1})}
\newcommand{\figu}[1]{\fig~\ref{fig:#1}}

\begin{document}
%%%%%%%%%%%%%%%%%%%%%%%%%%%%%%%%%%%%%%%%%%%%%%%%%%%%%%%%%%%%%%%%%%%%%
%%%%                     Title-page                              %%%%
%%%%%%%%%%%%%%%%%%%%%%%%%%%%%%%%%%%%%%%%%%%%%%%%%%%%%%%%%%%%%%%%%%%%%

\begin{titlepage}

% the footnote symbols are only redefined for the title page !
\renewcommand{\thefootnote}{\alph{footnote}}

\vspace*{-3.cm}
\begin{flushright}
TUM-HEP-656/07\\
MADPH-07-1478\\
\end{flushright}

\vspace*{0.5cm}

\renewcommand{\thefootnote}{\fnsymbol{footnote}}
\setcounter{footnote}{-1}

{\begin{center}
{\Large\bf New features in the simulation of neutrino oscillation experiments with GLoBES 3.0}

\end{center}}
{\begin{center}
{\large\bf (General Long Baseline Experiment Simulator)}
\end{center}}
\renewcommand{\thefootnote}{\alph{footnote}}

\vspace*{.8cm}
%\vspace*{.3cm}
{\begin{center} {\large{\sc
                Patrick~Huber\footnotemark[1],~
                Joachim~Kopp\footnotemark[2],~
                Manfred~Lindner\footnotemark[2], \\
                Mark~Rolinec\footnotemark[3],~
                Walter~Winter\footnotemark[4]
                }}

\footnote{All correspondence should be addressed to {\tt globes@mpi-hd.mpg.de}}

\end{center}}
\vspace*{0cm}
{\it
\begin{center}

\footnotemark[1]%${}^,$\footnotemark[2]%
       Department of Physics, University of Wisconsin, \\
       1150 University Avenue, Madison, WI 53706, USA

\vspace*{1mm}

\footnotemark[2]%
       Max--Planck--Institut f\"ur Kernphysik,  \\
       Postfach 10~39~80, D--69029 Heidelberg, Germany 

\vspace*{1mm}

\footnotemark[3]%
       Physik--Department, Technische Universit\"at M\"unchen, \\
       James--Franck--Strasse, 85748 Garching, Germany

\vspace*{1mm}

\footnotemark[4]%
       Universit\"at W\"urzburg, 
       Institut f\"ur theoretische Physik und Astrophysik, \\
       Am Hubland, D-97074 W\"urzburg, Germany

\end{center}}

\vspace*{1cm}

\begin{abstract}
We present Version 3.0 of the GLoBES (``General Long Baseline Experiment Simulator'') software, which is
a simulation tool for short- and long-baseline neutrino oscillation experiments. As a new feature, \GLOBESN\ allows for user-defined systematical errors, which can also be used to simulate experiments with
multiple discrete sources and detectors.
In addition, the combination with external information, such as from different experiment classes, is 
simplified. As far as the probability calculation is concerned, \GLOBES\ 
now provides an interface for the inclusion of non-standard physics without re-compilation of the software. 
The set of experiment prototypes coming with \GLOBES\ has been updated. For example, built-in fluxes are now
provided for the simulation of beta beams.
\end{abstract}

\vspace*{.5cm}

\begin{center}
\end{center}

\end{titlepage}

\newpage

\renewcommand{\thefootnote}{\arabic{footnote}}
\setcounter{footnote}{0}

\section{Introduction}

Neutrino oscillations are now established as the leading flavor
transition mechanism for neutrinos in a long history of many experiments, see
\eg\ \Ref~\cite{Barger:2003qi} and references therein. Future facilities, using accelerator-based
 neutrino beams or nuclear reactors as neutrino sources, are proposed
for precision measurements of the neutrino oscillation parameters. In the
simulation of these experiments, 
the presence of multiple solutions which are intrinsic to
 neutrino oscillation probabilities~\cite{Fogli:1996pv,Burguet-Castell:2001ez,Minakata:2001qm,Barger:2001yr} 
affect the performance.
Thus, optimization strategies are required which maximally exploit 
complementarity between experiments. The \GLOBES\ software package~\cite{Huber:2004ka}
is a  modern experiment  simulation  and analysis tool for
a highly accurate beam and detector simulation. In addition, it provides powerful 
means to analyze correlations and degeneracies, especially for the combination
of several experiments. Compared to a Monte Carlo simulation, which yields a different
result in each run, it simulates the performance of the {\em average} experiment
(see \Ref~\cite{Schwetz:2006md} for a discussion of the meaning of ``average'').
The advantage of such an average prediction is a tremendous performance gain,
which can be used for systematical parameter space scans. In addition, it 
simplifies the direct comparison of experiments.

The \GLOBES\ software has, in the past, been used for many studies, some of them
are referred to in the paper. However, recent developments have indicated that extensions
and improvements are necessary. In this work, we present the most important new
features and changes in \GLOBESN . For experimentalists, \GLOBES\ now allows
for the implementation of arbitrary systematics, which can also be used for
the simulation of multi-detector experiments. In addition, several extensions
have been included in \AEDL\ (``Abstract Experiment Description Language''), such as
built-in beta beam fluxes and the possibility to use lists as variables.
For phenomenologists, user-defined
priors provide a flexible interface to include external information, such as from
different experiments. In addition, \GLOBES\ now can be used for the simulation
of non-standard physics without re-compiling the \GLOBES\ software.

\section{Concept of \GLOBES }

\begin{figure}[t]
\begin{center}
\includegraphics[width=14cm]{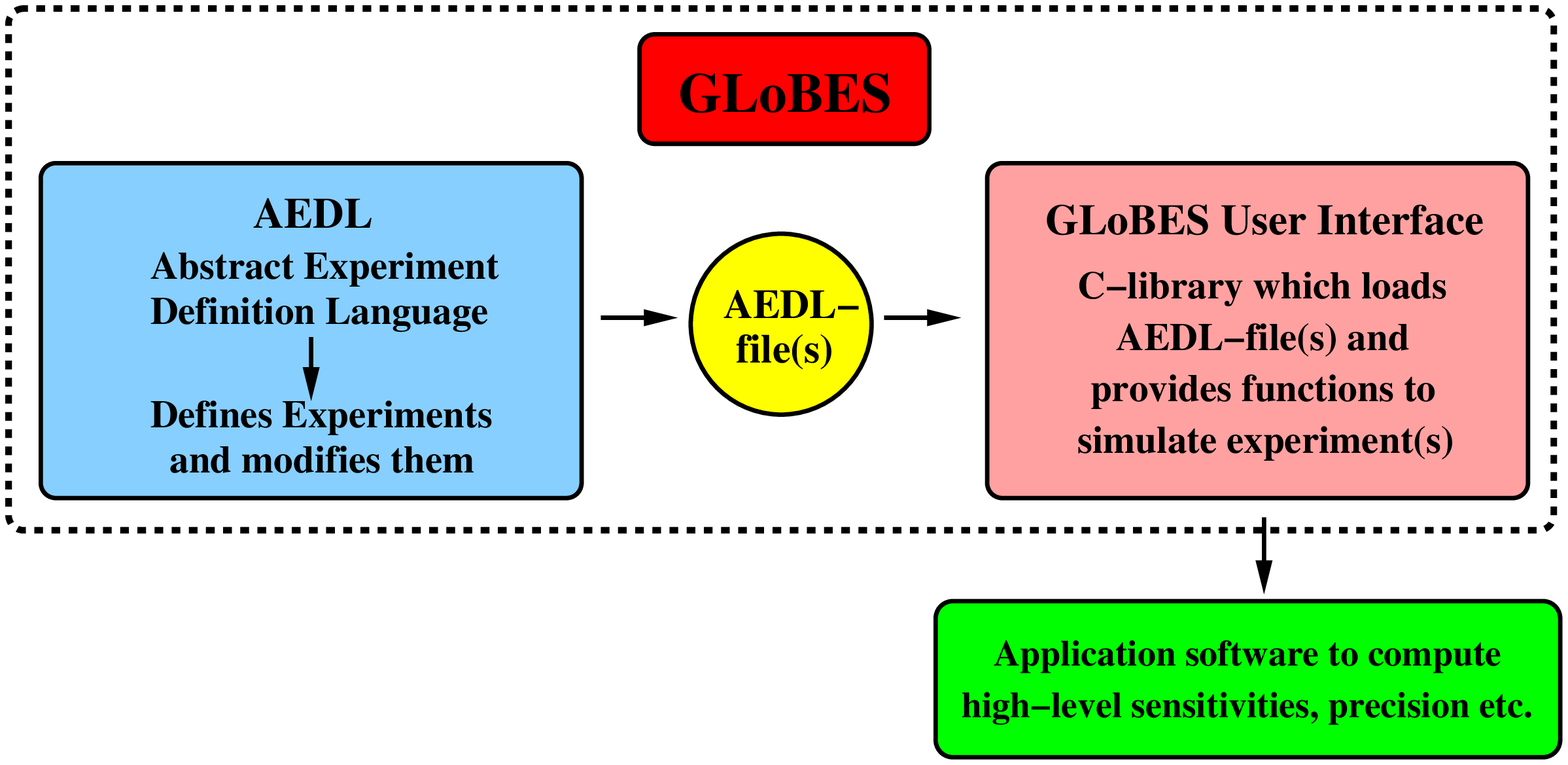}
\end{center}
\mycaption{\label{fig:GLOBES} General concept of the \GLOBES\ package.}
\end{figure}

\GLOBES\ (``General Long Baseline Experiment Simulator'') is a flexible
software tool to simulate and analyze neutrino oscillation 
short- and long-baseline experiments using a 
complete three-flavor description. On the
one hand, it contains a comprehensive abstract experiment definition
language (\AEDL ), which allows to describe 
most classes of long baseline and reactor experiments
at an abstract level. On the other hand, it provides a C-library to 
process the experiment information in order to obtain oscillation
probabilities, rate vectors, and $\Delta \chi^2$-values (\cf, \figu{GLOBES}). 
In addition, it provides a binary program to test experiment
definitions very quickly, before they are used by the application software.
Currently, \GLOBES\ is available for GNU/Linux. 

\GLOBES\ allows to simulate experiments with stationary neutrino point 
sources, where each experiment is assumed to have only one neutrino source.
Such experiments are neutrino beam experiments and reactor experiments. 
Geometrical effects of a continuous source distribution, such as in the sun or the 
atmosphere, can not be described. In addition, sources with a physically 
significant time dependence can not be studied, such as  supernov\ae. 
However, in \GLOBESN\ and higher, new flexibility is introduced by the
concept of user-defined systematics. This new feature allows the cross-definition
of systematical errors over different experiments. In principle, 
this mechanism can be used for the combination of several discrete sources
and one detector, or several detectors and one source, or several sources and
several detectors. In addition, already implemented
concepts of \GLOBES\ have been used for indirect simulations of geometrical
effects. For example, the mapping of the detector location on the neutrino
energy has been simulated in \Ref~\cite{Rolinec:2006xr} by the use of
variable bin widths.

On the experiment definition side, either built-in neutrino fluxes
(\eg, neutrino factory, beta beam) or arbitrary, user-defined fluxes can be used. 
Similarly,
arbitrary cross sections, energy dependent efficiencies,
energy resolution functions as well as the considered oscillation channels, 
backgrounds, and many other properties can be specified. 
For the systematics, energy
normalization and calibration errors can be simulated, or the
systematics can be completely user-defined. Note that
energy ranges and windows and bin widths can be
(almost) arbitrarily chosen, including bins of different
widths. Together with \GLOBES\ comes a number of
pre-defined experiments in order to demonstrate the capabilities
of \GLOBES\ and to provide prototypes for new experiments.
In addition, they can be used to test new physics ideas with
complete experiment simulations.

With the C-library, one can extract the $\Delta \chi^2$ for all defined 
oscillation channels for an experiment or any combination of experiments.
Of course, also low-level information, such as oscillation
probabilities or event rates, can be obtained. \GLOBES\ includes the
simulation of neutrino oscillations in matter with arbitrary matter 
density profiles. In addition, it allows to simulate the matter density
uncertainty (see, \eg, \Refs~\cite{Huber:2002mx,Ohlsson:2003ip}) and to 
extract the precision on the matter density (see, \eg, \Refs~\cite{Winter:2005we,Gandhi:2006gu}).
 As one of the most
advanced features of \GLOBES , it provides the technology to 
project the $\Delta \chi^2$, which is a function of all oscillation
parameters, onto any subspace of parameters by local minimization. 
This approach allows the inclusion of multi-parameter-correlations,
where external constraints (\eg, on the solar parameters) can be imposed, too.
Applications of the projection mechanism include the projections onto 
the $\stheta$-axis and the $\stheta$-$\deltacp$-plane. In addition, 
all oscillation parameters can be kept free to numerically localize 
degenerate solutions.

\section{Oscillation probabilities and the simulation of non-standard physics}

The probability calculation in \GLOBES\ is based on the diagonalization of the
Hamiltonian in layers of constant matter density using the standard three flavor
scenario of neutrino oscillations. In the flavor base, we have
\begin{equation}
\mathcal{H}(n_e) = \frac{1}{2E} \, U \,
\left(
\begin{array}{ccc}
0 & 0 & 0 \\
0  & \sdm & 0 \\
0  & 0 & \ldm \\
\end{array}
\right) \,
U^\dagger + 
\left(
\begin{array}{ccc}
 \pm \sqrt{2} G_F n_e & 0 & 0 \\
0 & 0 & 0 \\
0 & 0 & 0 
\end{array}
\right)
\label{equ:ham}
\end{equation}
with $U$ being the mixing matrix (described by four parameters $\theta_{12}$, $\theta_{13}$, $\theta_{23}$, and $\deltacp$), and $n_e$ the constant electron density
in the respective matter density layer. The electron density is related to the matter density $\rho$ by 
$n_e \simeq 0.5 \, \rho/m_N$ with $m_N$ being the nucleon mass. The sign in the second term depends on whether one
uses neutrinos or antineutrinos. For $k$ matter density layers with densities
$n^k_e$ and thicknesses $x_i$, the oscillation probability then evaluates to
\begin{equation}
P(\nu_\alpha \rightarrow \nu_\beta) = \left| \langle \nu_\beta | \mathcal{E}(n^k_e, x_k) \hdots  \mathcal{E}(n^1_e, x_1) | \nu_\alpha \rangle \right|^2
\label{equ:probs}
\end{equation}
with the evolution operators (see, \eg, \Ref~\cite{Ohlsson:1999um})
\begin{equation}
  \mathcal{E}(n^k_e, x_k) = \exp \left( -i \mathcal{H}(n^k_e) x_k \right) = \frac{1}{2E} \, \hat{U}(n^k_e) \, \left( \begin{array}{ccc}
e^{-i \, m_1^k \, x_k} & 0 & 0 \\
0 &  e^{-i \, m_2^k \, x_k} & 0 \\
0 & 0 & e^{-i \, m_3^k \, x_k} 
\end{array}
\right) \, \hat{U}(n^k_e)^\dagger \, .
\end{equation}
Here the Hamiltonian $\mathcal{H}(n^k_e)$ is diagonalized by the unitary mixing matrix in matter $\hat{U}(n^k_e)$
with the eigenvalues $m_1^k$, $m_2^k$, and $m_3^k$. The probabilities in \equ{probs} are used for the
event rate computation, which is described in greater detail in \Ref~\cite{Huber:2004ka}. In the next section,
we will demonstrate how systematics is implemented.

\begin{figure}[t]
\begin{center}
\includegraphics[width=8cm]{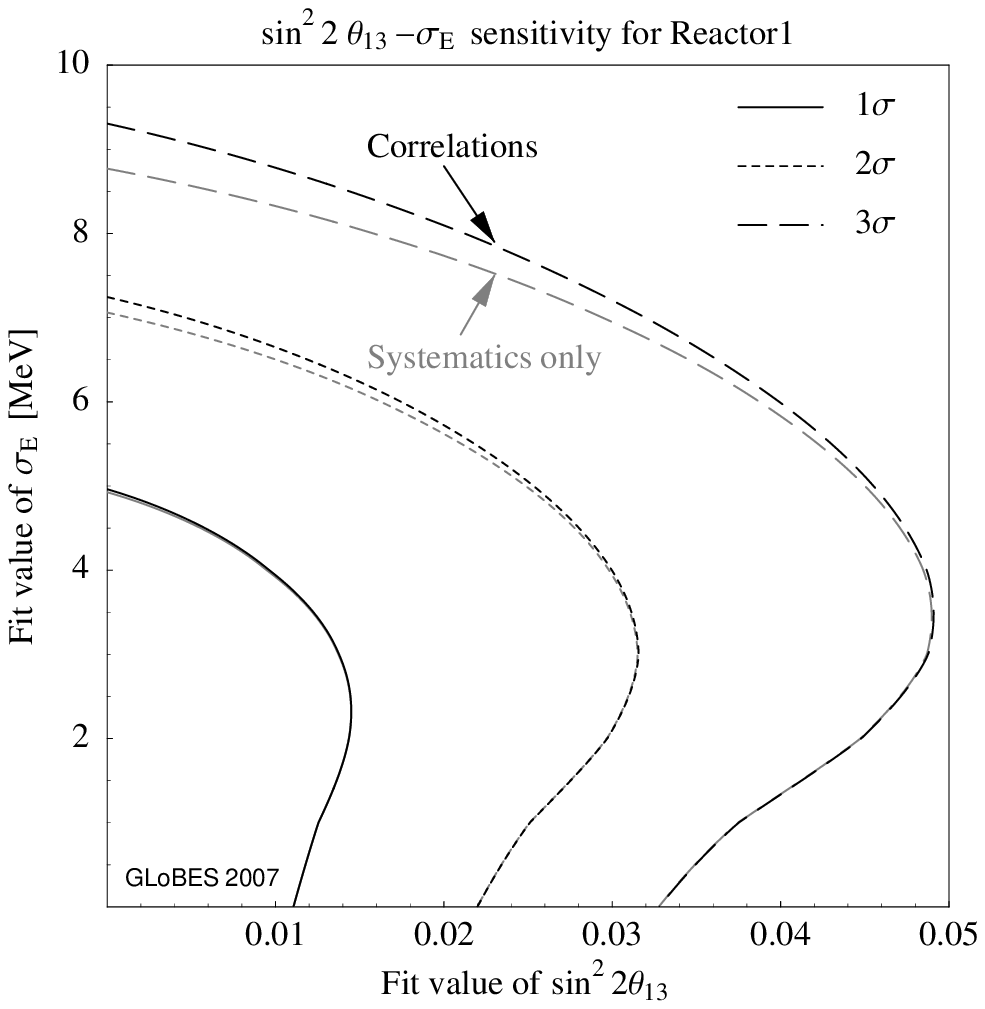}
\end{center}
\mycaption{\label{fig:reactorNSI} Fit in the $\stheta$-$\sigma_E$-plane for the simulated values
$\stheta=\sigma_E=0$ (figure similar to \Ref~\cite{Blennow:2005yk}). Here $\sigma_E$ is an additional non-standard parameter describing the wave packet width of the neutrino oscillation wave packet. If this parameter is
kept free (vertical axis), the $\stheta$ sensitivity (horizontal axis) will be affected. Note that
the standard $\stheta$ sensitivity is given by the curves for $\sigma_E \equiv 0$.
The figure is computed with
{\tt example6.c} coming with the \GLOBESN\ distribution (and higher).}
\end{figure}

\GLOBESN\ allows the modification of this Hamiltonian and the whole probability engine without
re-compilation of the \GLOBES\ software. This is implemented
by using pointers to the functions of the standard probability engine, which can be changed by
registering a different probability engine. For example, a user may just copy the standard probability
engine of \GLOBES , modify it, and register it. An important prerequisite is the ability to handle
more than the standard six oscillation parameters (plus matter density) $\boldsymbol{\lambda}=(\theta_{12},\theta_{13}, \theta_{23}, \deltacp, \sdm, \ldm, \rho)$, \ie,
\begin{equation}
\boldsymbol{\lambda}=(\theta_{12},\theta_{13}, \theta_{23}, \deltacp, \sdm, \ldm, \rho, \eta_1, \hdots, \eta_n)
\end{equation}
with $n$ non-standard parameters $\eta_j$. These parameters can be accessed in the usual way, see \Ref~\cite{Manual}.
For example, non-standard Hamiltonian effects in the $e$-$\tau$-sector may be introduced by modifying \equ{ham} to
\begin{equation}
 \mathcal{\hat{H}} = \mathcal{H}(n_e) + 
\left(
\begin{array}{ccc}
0 & 0 & \epsilon_{e \tau} \\
0 & 0 & 0 \\
\epsilon_{e \tau}^* & 0 & 0 
\end{array}
\right)
\label{equ:nsi}
\end{equation}
In this case, there are two more real parameters, such as the absolute value and phase of $\epsilon_{e \tau}$.
In the literature, the non-standard physics feature in \GLOBES\ as experimental feature has been used for damping effects (such as neutrino decoherence, decay, \etc) in \Ref~\cite{Blennow:2005yk}, Hamiltonian-level effects (such as non-standard matter effects) in \Ref~\cite{Blennow:2005qj}, and mass-varying neutrinos in \Ref~\cite{Schwetz:2005fy} by implementing an environment dependence of neutrino mass. 
We show in \figu{reactorNSI} a result from \Ref~\cite{Blennow:2005yk} provided as {\tt example6.c} in the \GLOBES\
distribution. In this case, the non-standard physics is a loss of coherence in neutrino oscillations
described by an intrinsic width of the mass eigenstate wave packets $\sigma_E$ (in energy space).
Compared to \equ{nsi}, the underlying physics takes place at the probability level, \ie, 
\equ{probs}, and it can be easily implemented analytically (for details, see \Ref~\cite{Blennow:2005yk}). 
In {\tt example6.c}, an analytical implementation
of this physics is used for simplicity.

\section{Systematics implementation and user-defined systematics}

\GLOBES\ supports four types of systematical errors by default: Signal and background
normalization errors, as well as signal and background tilt or energy calibration errors.
The tilt (T) is implemented as a linear distortion of the spectrum around the center,
and the energy calibration (C) as a distortion (stretching) of the reconstructed energy scale.
For the implementation of these systematical errors, the  ``pull method'' is used~\cite{Fogli:2002pt}, which introduces  so-called nuisance parameters $\zeta_i$. 
For example, for the signal and background normalization errors, the respective event rates 
$s_i$ and $b_i$ in
each bin $i$ are multiplied with\footnote{For details on the event rate calculation, see \Ref~\cite{Huber:2004ka}.}
\begin{equation}
s_i(\zeta_1) := (1+\zeta_1)\cdot s_i \, , \quad b_i(\zeta_2) : =(1+\zeta_2)\cdot b_i \, ,
\end{equation} 
\ie, the rates are scaled by these parameters. The systematics $\chi^2$ is then minimized over
these parameters:
\begin{equation}
\chi^2_\mathrm{pull}(\boldsymbol{\lambda}):=\min_{\{\zeta_i\} } \,\, \left( \,
\chi^2(\boldsymbol{\lambda},
\zeta_1, \ldots, \zeta_k)+ \sum_{j=1}^{k} \frac{\zeta_j^2}{\sigma_{\zeta_j}^2}
\, \right)\,.
\label{equ:chipull}
\end{equation}
Here $\chi^2(\boldsymbol{\lambda}, \zeta_1, \ldots, \zeta_k)$ is the usual Possonian $\chi^2$
depending on the neutrino oscillation parameters $\boldsymbol{\lambda}$ and the nuisance
parameters $\zeta_i$ (which, for instance, scale the signal or background rates).
In addition, Gaussian penalties $\zeta_j^2/\sigma_{\zeta_j}^2$ are added, where 
$\sigma_{\zeta_j}$ corresponds to the actual systematical error.
The core part of \equ{chipull} is the definition of $\chi^2$. For example, for a background-free measurement and
a signal normalization error only, it reads
\begin{equation}
  \chi^2 (\boldsymbol{\lambda},a) =  \sum_{i=1}^{\textrm{\# of bins}}2
     \left( (1 + a) T_{i} - O_{i} +
       O_{i} \log \frac{O_{i}}{(1 + a) T_{i}}  \right) + \frac{a^2}{\sigma_a^2} \, ,
\label{equ:chistandard}
\end{equation}
where $O_i$ are the observed rates (corresponding to the data or true values), and $T_i$ are
the theoretical (fit) rates.
Let us assume that we have two detectors, such as for a reactor experiment with a near and
far detector. Using standard systematics in \GLOBES , one has
\begin{equation}
\chi^2_\mathrm{pull}(\boldsymbol{\lambda}) = \min_{a } \,\, \left( \,
\chi_1^2(\boldsymbol{\lambda},a)+ \frac{a^2}{\sigma_{a}^2} \, \right)
+\min_{ b } \,\, \left( \,
\chi_2^2(\boldsymbol{\lambda},b)+ \frac{b^2}{\sigma_{b}^2}
\, \right)\,.
\label{equ:chipullex}
\end{equation}
In this case, the normalization between the two detectors will be uncorrelated.

\begin{figure}[t]
\begin{center}
\includegraphics[width=9cm]{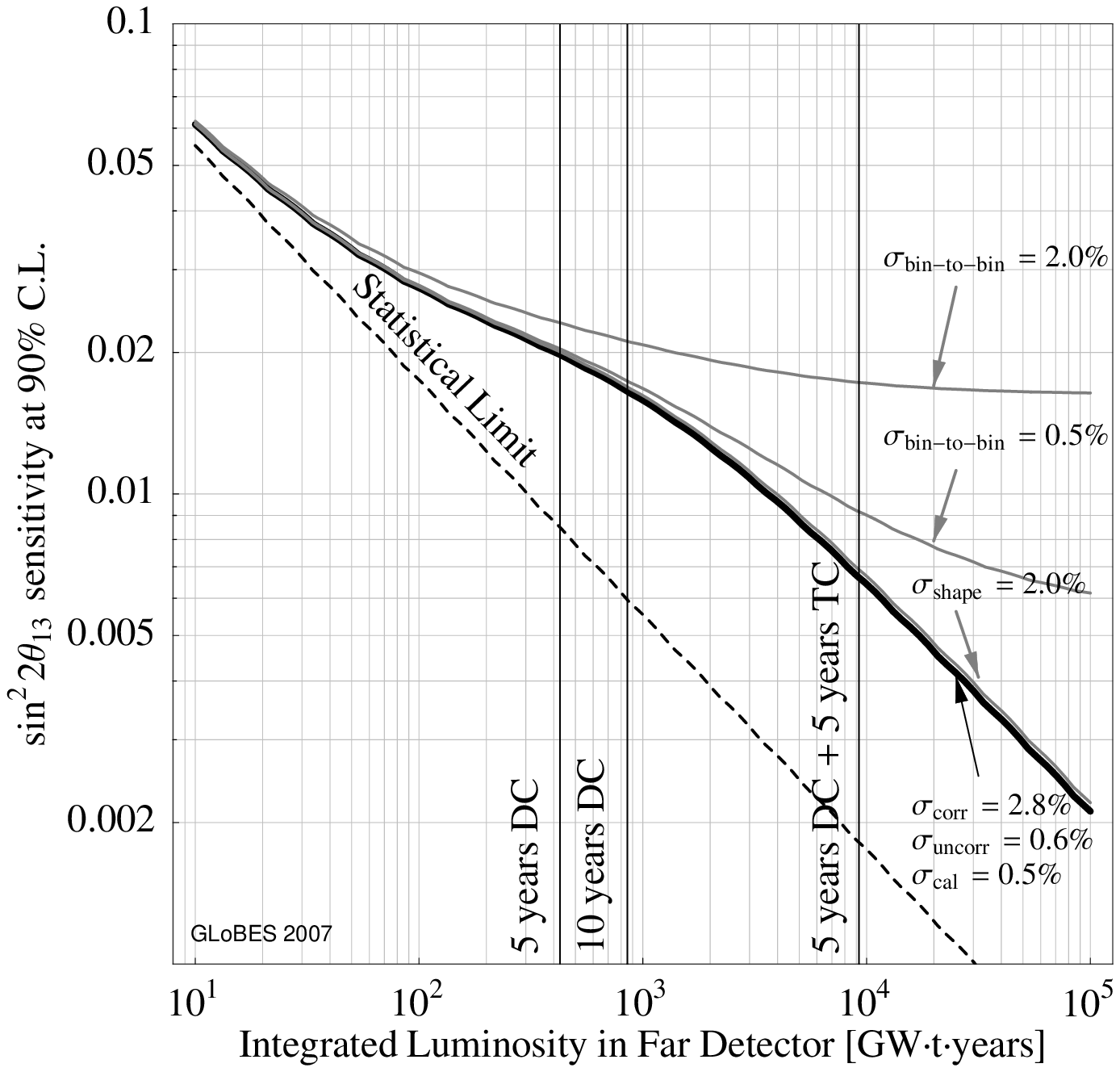}
\end{center}
\mycaption{\label{fig:reactor} Luminosity scaling of the $\stheta$ sensitivity for different systematical error assumptions (90\% confidence level, figure similar to \Ref~\cite{Huber:2006vr}). The thick curve corresponds to \equ{chi2reactor} for $\sigma_R=\sigma_{\mathrm{corr}}=2.8\%$, $\sigma_N=\sigma_F=\sigma_{\mathrm{uncorr}}=0.6\%$ and an additional calibration error. The figure is computed with {\tt example5.c} provided with the \GLOBES\ software.}
\end{figure}

Using \GLOBESN\ and higher, one can modify the standard $\chi^2$ function in
\GLOBES\ which corresponds to \equ{chistandard}. In addition, the systematics
concept of ``error dimensions'' is replaced by a concept directly related to
this $\chi^2$ function. As an example, consider a reactor experiment with near ($N$)
and far ($F$) detectors which are described by a reactor flux uncertainty $\sigma_R$
and two fiducial volume errors of the near $\sigma_N$ and far $\sigma_F$ detectors.
Then \equ{chistandard} will be replaced by
\begin{align}
  \chi^2(\boldsymbol{\lambda},a_N,a_F,a_R) = & \sum_{i=1}^{\textrm{\# of bins}} \sum_{d = N,F} 2
     \Big( (1 + a_R + a_d) T_{d,i} - O_{d,i} +  \nonumber \\
& \left. + O_{d,i} \log \frac{O_{d,i}}{(1 + a_R + a_d) T_{d,i}}  \right) 
          + \frac{a_R^2}{\sigma_R^2} + \frac{a_N^2}{\sigma_N^2} + \frac{a_F^2}{\sigma_F^2}
\label{equ:chi2reactor}
\end{align}
This function can be defined with user-defined systematics. Similarly, calibration errors, shape
errors, uncorrelated bin-to-bin errors, \etc , can be introduced. We show in \figu{reactor} the result of
such a calculation, where the thick curve corresponds to \equ{chi2reactor} plus an additional 
energy calibration error.\footnote{In fact, \figu{reactor} was computed with Gaussian statistics for
efficiency reasons, whereas \GLOBES\ internally uses Poissonian statistics as a standard; see \Ref~\cite{Manual}.}
Furthermore, more complicated setups using multiple discrete sources and detectors can be simulated. For details, we refer to the manual~\cite{Manual}.
In the past, user-defined systematics has (as experimental feature) been used in \Ref~\cite{Huber:2006vr} to
simulate Double Chooz, in \Ref~\cite{Barger:2006kp} to simulate T2KK, and in
\Ref~\cite{Kopp:2006mw} to simulate various reactor experiment setups and complicated
 backgrounds from geoneutrinos. 

In summary, \GLOBES\ currently supports the following systematics $\chi^2$ functions:
\begin{description}
\item[{\tt chiSpectrumTilt}]
 Spectral information, signal and background normalization errors and tilts 
\item[{\tt chiNoSysSpectrum}]
 No systematical errors, but spectral information
\item[{\tt chiTotalRatesTilt}]
 Total rates only, signal and background normalization errors and tilts 
\item[{\tt chiSpectrumOnly}]
 Spectrum only (free signal normalization), no other systematics 
\item[{\tt chiNoSysTotalRates}]
 Total rates only, no systematics
\item[{\tt chiSpectrumCalib}]
 Spectral information, signal and background normalization errors and energy calibration errors 
\item[{\tt chiZero}]
 Passive systematics (returns $\chi^2=0$, only provides the event rates for the access by other
rules or experiments)
\item[{\tt Any other name}]
 User-defined systematics
\end{description}

\section{Adding external information with user-defined priors}

In order to include correlations and degeneracies, \GLOBES\ provides
(local) marginalization routines to project the fit manifold onto
a subspace. For example, if one wants to compute the $\stheta$ precision,
one scans the $\stheta$ direction and marginalizes for each fixed $\stheta$
over the other oscillation parameters. This marginalization is performed
after the systematics $\chi^2_{\mathrm{pull}}$ in \equ{chipull} (or similar) 
has been determined:\footnote{The marginalization order systematics - correlations
is, strictly speaking, not valid anymore for the new hybrid minimizer provided
with \GLOBESN\ as an experimental feature. However, we keep this description
for pedagogical reasons.}
\begin{equation}
 \chi^2_{\mathrm{proj}}(\boldsymbol{\lambda_1}) := \min_{\{ \boldsymbol{\lambda_2} \} } \left( \, \chi^2_{\mathrm{pull}} (\boldsymbol{\lambda_1} , \boldsymbol{\lambda_2} ) \, \right) \, .
\label{equ:margin}
\end{equation}
In this case, the parameter space of oscillation parameters 
$\boldsymbol{\lambda}$ is projected onto the subspace $\boldsymbol{\lambda_1}$.
If one wants to include external input, which only depends on the oscillation
parameters and not on systematics, one can do so at this level. \GLOBES\ provides
the pre-defined possibility to include external Gaussian constraints on the 
oscillation parameters (and the matter density, which is treated as another
oscillation parameter). For example, an external constraint on $\theta_{12}$
is added before the marginalization in \equ{margin} by the replacement
\begin{equation}
 \chi^2_{\mathrm{pull}}(\boldsymbol{\lambda}) \rightarrow \chi^2_{\mathrm{pull}}(\boldsymbol{\lambda})  +
 \frac{(\theta_{12} - \theta_{12}^0)^2}{\sigma_{\theta_{12}}^2}
\label{equ:priors}
\end{equation}
with the central value $\theta_{12}^0$, where the input is added, and
the $1\sigma$ Gaussian error $\sigma_{\theta_{12}}$. This corresponds to an external
measurement of the solar mixing angle at the best-fit value $\theta_{12}^0$
with the error $\sigma_{\theta_{12}}$. 

In practice, the external error on $\theta_{12}$ may not be Gaussian or one may
want to add other external information. Therefore,
\GLOBESN\ and higher provides the concept of user-defined priors  $f(\boldsymbol{\lambda})$. 
Instead of \equ{priors}, we then have
\begin{equation}
 \chi^2_{\mathrm{pull}}(\boldsymbol{\lambda}) \rightarrow \chi^2_{\mathrm{pull}}(\boldsymbol{\lambda})  + f(\boldsymbol{\lambda}) \, .
\label{equ:userpriors}
\end{equation}
One can easily imagine that this simple concept allows for a high degree of freedom. Examples where
this feature has been used as an experimental feature are the combination of terrestrial neutrino
data with the information from neutrino telescopes~\cite{Winter:2006ce} and the combination of long-baseline data with atmospheric neutrino data~\cite{Huber:2005ep,Campagne:2006yx}.
Another very interesting option is the use of penalties in degeneracy localization, as suggested by Thomas
Schwetz. For example, if the minimizer runs into the wrong octant, user-defined priors can be
used to add a penalty and prevent it from doing so.

\section{\AEDL\ changes and experiment prototypes}

\begin{table}[t!]
{\small
\begin{center}
\begin{tabular}{llp{7.0cm}c}
\hline
Experiment & File name & Short description &  \Refs \\
\hline 
\multicolumn{4}{l}{\underline{Superbeam experiments:}} \\
T2K & {\tt T2K.glb} & J-PARC to Super-Kamiokande &  \cite{Itow:2001ee,Huber:2002mx} \\
T2HK & {\tt T2HK.glb} & J-PARC to Hyper-Kamiokande &  \cite{Itow:2001ee,Huber:2002mx} \\
NO$\nu$A & {\tt NOvA.glb} & Fermilab NuMI beamline off-axis  & \cite{Ambats:2004js,Yang_2004} \\
SPL & {\tt SPL.glb} & CERN to Fr\'{e}jus & \cite{Campagne:2006yx,Campagne:2004wt,Mezzetto:2003mm} \\[0.2cm]
\multicolumn{4}{l}{\underline{Reactor experiments:}} \\
{\sc Reactor-I} & {\tt Reactor1.glb} & Small reactor exp., 
$\mathcal{L} = 400 \, \mathrm{t} \, \mathrm{GW} \, \mathrm{yr}$
& \cite{Huber:2003pm} \\
{\sc Reactor-II} & {\tt Reactor2.glb} & Large reactor exp., 
$\mathcal{L} = 8\, 000 \, \mathrm{t} \, \mathrm{GW} \, \mathrm{yr}$ 
& \cite{Huber:2003pm} \\
Double{\sc Chooz} & {\tt D-Chooz\_near.glb} & Double Chooz near 
detector & \cite{Huber:2006vr} \\ 
 & {\tt D-Chooz\_far.glb} & Double Chooz far detector & \\[0.2cm]
\multicolumn{4}{l}{\underline{Beta beams:}} \\
Low $\gamma$ & {\tt BB\_100.glb} & $\gamma=100$ CERN to Fr\'{e}jus 
scenario & \cite{Campagne:2006yx} \\
Medium $\gamma$ & {\tt BB\_350.glb} & $\gamma=350$ ``refurbished SPS'' scenario (or other accelerator)
& \cite{Burguet-Castell:2005pa} \\
Variable $\gamma$ & {\tt BBvar\_WC.glb} & Variable $\gamma$ beta beam with water Cherenkov detector, $50 \lesssim \gamma \lesssim 500$ & \cite{Huber:2005jk} \\[0.2cm]
	& {\tt BBvar\_TASD.glb} & Variable $\gamma$ beta beam with totally active scint. detector (TASD), 
$100 \lesssim \gamma \lesssim 3\, 000$ & \cite{Huber:2005jk} \\[0.4cm]
\multicolumn{4}{l}{\underline{Neutrino factories:}} \\
Standard & {\tt NFstandard.glb} & Standard neutrino factory, $50 \, \mathrm{kt}$ magnetized iron calorimeter, 
$E_\mu = 50 \, \mathrm{GeV}$ & \cite{Huber:2002mx} \\
Variable $E_\mu$ & {\tt NFvar.glb} & Variable neutrino factory, disapp. channels
{\em without} CID; $10 \, \mathrm{GeV} \lesssim E_\mu \lesssim 80 \, \mathrm{GeV}$ & \cite{Huber:2002mx,Huber:2006wb}  \\
Gold + Silver & {\tt NF\_GoldSilver.glb} & As {\tt NFvar.glb} 
plus 5~kt ECC detector for Silver Channel
measurement & \cite{Huber:2002mx,Huber:2006wb,Autiero:2003fu} \\
Hybrid detector & {\tt NF\_hR\_lT.glb} & As {\tt NFvar.glb}, but 
lower threshold and better energy resolution & \cite{Huber:2002mx,Huber:2006wb} \\
\hline
\end{tabular}
\end{center}
} % small
\mycaption{\label{tab:experiments} 
Pre-defined experiment prototypes coming with \GLOBESN : The columns represent the \AEDL\ filename, 
a short description, and the references in which the files are originally used and 
discussed (except for minor modifications, such as a different implementation of 
the energy threshold function). 
}
\end{table}

\AEDL\ (Abstract Experiment Definition Language) is a special syntax developed for the description of
experiments in \GLOBES, represented by simple text files. Core part of \AEDL\ are the following constructions
(for details, see \Ref~\cite{Manual}):
\begin{description}
\item[Channel] A channel represents an oscillation channel from the source flux, over the initial and final
flavors and the use of neutrinos or antineutrinos, to the interaction type and cross section. The result
is an event rate vector.
\item[Rule] A rule combines one or more signal channels with one or more background channels. The rates
from these channels are added, and the signal and background are implemented with specific systematics, \ie,
systematics is connected to the concept of a rule.
The result of a rule is a systematics $\chi^2$ corresponding to \equ{chipull}.
\item[Experiment] An experiment is a combination of one or more rules, which may correspond to different
appearance and disappearance channels, neutrino and antineutrino operations modes, \etc. An experiment has a common detector and baseline, but may use several different fluxes (such as for neutrinos and antineutrinos). Therefore,
the matter density profile is the same for all rules of an experiment. The result of an experiment
is again a systematics $\chi^2$, but summed over all rules.
\item[Experiment combination] An experiment combination consists of different experiments with different detectors, baselines, and (uncorrelated) matter density profiles. The result of an experiment combination
is again a systematics $\chi^2$, but summed over all experiments.
\end{description}
Note that after the $\chi^2$ is summed over all rules or experiments as specified by the user in the application software, \equ{margin} is applied, and the oscillation parameters are marginalized over. 

In \GLOBESN , this construction of experiments is extended by user-defined systematics, which allows
the user to break up the strict coupling between systematics and rules. For example, systematics can be
correlated between different rules or experiments, such as for a reactor experiment with identical
near and far detectors. In addition, one can introduce user-defined systematical errors.
For that purpose, the former error dimension concept has been changed and
extended. For example, one has to give each user-defined systematics a name, and this name has
to be matched by the application software. 
Further changes in \AEDL\ are the requirement to specify the minimum \GLOBES\ version the \AEDL\ file can be used with,
and the extensions by lists as \AEDL\ variables and by an interpolation feature.

As far as the experiment prototypes are concerned, a list of the ones provided with \GLOBESN\ 
is given in \Tab~\ref{tab:experiments}. These prototypes can be modified for the definition of
one's own experiments (experimentalist), or they can be used for the test of new physics concepts
(phenomenologist/theorist).
Note that some of these experiments may contain 
integrated luminosities, baseline, fluxes, efficiencies, or other factors, which the user
may not agree with and which can be easily adjusted. Some of the \AEDL\ files have been
changed or updated from earlier versions, such as for T2K and T2HK, or the standard neutrino
factory. Other files represent the results from recent developments, such as the variable energy
beta beam or neutrino factory files. They push \AEDL\ to the edge in terms of advanced features,
such as variable bin widths, use of the new interpolation feature, threshold function implementations, \etc.
In particular, note that Double Chooz requires user-defined
systematics, and the beta beams require built-in beta beam fluxes. Both of these features are only
available in \GLOBESN\ and higher. Some of the files require that \AEDL\ variables be pre-set by the
user, such as the muon energy for the variable energy neutrino factory. These variables are
described in the corresponding file headers. Old \AEDL\ prototypes should still run, but will not
be maintained by the \GLOBES\ team anymore.

\section{Internal changes and improved performance}

Sophisticated \GLOBES\ simulations with several experiments and complicated
parameter correlations can take several days to finish even on modern computers.
To mitigate this problem as far as possible, we have taken several steps to
optimize the performance of the multi-dimensional minimizations, which are at
the heart of each \GLOBES\ $\chi^2$ analysis.

Firstly, the \GLOBES\ minimizers distinguish between minimization over
oscillation parameters and minimization over systematics (nuisance) parameters.
This is necessary because each step in the oscillation parameter space
requires a time-consuming re-computation of oscillation probabilities,
while a modification of the systematics parameters does not. The standard
procedure is to perform a full minimization in systematics space after each
step in oscillation space. This avoids interference of the two categories
of parameters, and is known to have excellent convergence properties.
The numerical method that does the minimization is the Powell
algorithm~\cite{Fletcher:1963:RCD}.

In \GLOBESN , we provide an alternative method, which does the full minimization
in one go, and is therefore several times faster in most cases. In this algorithm,
each Powell iteration over oscillation parameters is followed by one iteration
over systematics parameters. Since there may be some pathological cases where this
interleaved minimization strategy behaves differently than the old method,
it is by default not used in \GLOBESN . This ensures maximal compatibility
with old application programs, but it requires users who wish to benefit from the
new algorithm to explicitly select it by an API function call.

Besides optimizing the minimization functions, it is also mandatory to
optimize the routines which are called by them to calculate $\chi^2$
for one specific set of parameters. In particular, the computation of
oscillation probabilities, which has to be done every time the oscillation
parameters have changed, should be as efficient as possible. The most expensive
step here is the diagonalization of the hermitian $3\times3$ neutrino Hamilton
operator in matter. The LAPACK routines~\cite{Anderson:LAPACK}, which were
used to solve this eigenproblem in previous versions of \GLOBES, turned out
not to be the optimal choice because they are mainly optimized for very large
matrices~\cite{Kopp:2006wp}. In the case of the small and usually well-conditioned
matrices appearing in \GLOBES, they spend too much time evaluating their parameter
list and planning the optimal diagonalization strategy. Therefore, \GLOBES\ now
relies an algorithm which has been designed specifically for the diagonalization
of hermitian $3\times3$ matrices. It is essentially the extremely fast analytical
algorithm discussed in \Ref~\cite{Kopp:2006wp}, but for the rare cases where this may become
inaccurate, it falls back to the QL method~\cite{Fra61,Stoer:NumAnalysis}.

As far as the installation of \GLOBES\ on different systems is concerned, \GLOBES\
has been improved in the past, such as to handle different Linux distributions.
In addition, for high-throughput computing, \GLOBES\ has been used or tested
on 64 Bit systems, Condor clusters, and parallel clusters. An important prerequisite
for the parallelization is the option to build static binaries, which is now
illustrated in the examples. With these improvements, the performance of \GLOBES\
has been pushed to the edge. \Ref~\cite{Huber:2006wb} is probably the currently
most demanding product of \GLOBES , for which most of these features
have been developed.

\section{Summary and conclusions}

We have presented \GLOBESN , which is a simulation software for 
future terrestrial neutrino oscillation facilities. \GLOBES\ is now no further
restricted to one source-one detector configurations, but it can simulate
the systematics of arbitrary combinations of discrete sets of sources and
detectors. In addition, user-defined systematical errors can be introduced.
The new concept of user-defined priors allows the inclusion of external
information at a different level. For example, constraints from atmospheric
neutrino experiments or neutrino telescopes can be taken into account. In addition,
they provide a powerful handle on complicated degeneracy localization.
On the probability level, \GLOBES\ now supports the simulation of non-standard
physics beyond three-flavor neutrino oscillations without a re-compilation of
the software. A very important prerequisite for this feature is the ability
of \GLOBESN\ to carry more than six oscillation parameters. New experiment
prototypes make extensive use of these new features, as well as they are illustrated by
examples in the manual. One prominent example is the Double Chooz experiment,
which requires user-defined systematics to simulate correlated errors between
the near and far detectors.

In conclusion, \GLOBESN\ provides a number of new improvements, such as by
 including more flexibility. At the end, the developers of a software can never
exactly foresee what the users are up for, and therefore flexible concepts
allow for more space in problem realizations. In this version of \GLOBES ,
three new such flexible concepts have been introduced: user-defined oscillation 
probabilities, user-defined systematics, and user-defined priors. These approaches
will make an easy adjustment of the software possible for both experimentalists and theorists,
and will allow for flexibility at all levels.
Finally, note that the past has shown that there is often more than one possibility 
how a specific problem can be implemented in \GLOBES .

\subsection*{Acknowledgments}

For the current version of \GLOBES , we would like to 
thank Thomas Schwetz, who has been pushing the software to the edge in the 
past few years. Furthermore, we would like to thank Tommy Ohlsson,
Toshihiko Ota, and Julian Skrotzki for using and testing unpublished new features of the software.
The development of \GLOBES\ is currently being supported by the
Emmy Noether-Program of Deutsche Forschungsgemeinschaft [WW].
In the past, the following institutions and organizations have supported 
members of the \GLOBES\ team: Technische Universit\"at M\"unchen,
Max-Planck-Institut f\"ur Physik, SFB 375 for astro-particle physics
of Deutsche Forschungsgemeinschaft, Studienstiftung des Deutschen Volkes,
Institute for Advanced Study, Princeton, W. M. Keck Foundation, and National Science Foundation.

\end{document}